\documentclass[conference]{IEEEtran}
\usepackage{amsmath,amssymb,amsfonts}
\usepackage{algorithmic}
\usepackage{graphicx}
\usepackage{textcomp}
\usepackage{xcolor}
\usepackage{booktabs} 
 
\usepackage[british]{babel}
\usepackage{comment}
\usepackage{orcidlink}
\usepackage{csquotes}
\usepackage{calc}

\usepackage[acronym]{glossaries} 
\newacronym{api}{API}{application programming interface}
\newacronym{oss}{OSS}{open-source software}

\hypersetup{
colorlinks=true,
linkcolor=black,
citecolor=black,     
urlcolor=black
}
 
\begin{document}

\title{An Exploratory Study of Documentation Strategies for Product Features in Popular GitHub Projects
}

\author{
\IEEEauthorblockN{Tim Puhlfürß \orcidlink{0000-0001-8421-8071}, Lloyd Montgomery \orcidlink{0000-0002-8249-1418}, Walid Maalej \orcidlink{0000-0002-6899-4393}}
\IEEEauthorblockA{\textit{Universität Hamburg, Department of Informatics} \\  
Hamburg, Germany \\
tim.puhlfuerss@uni-hamburg.de, lloyd.montgomery@uni-hamburg.de, walid.maalej@uni-hamburg.de
}
}

\maketitle

\begin{abstract}
[Background]
In large open-source software projects, development knowledge is often fragmented across multiple artefacts and contributors such that individual stakeholders are generally unaware of the full breadth of the product features.
However, users want to know what the software is capable of, while contributors need to know where to fix, update, and add features.
[Objective]
This work aims at understanding how feature knowledge is documented in GitHub projects and how it is linked (if at all) to the source code.
[Method]
We conducted an in-depth qualitative exploratory content analysis of 25 popular GitHub repositories that provided the documentation artefacts recommended by  \textit{GitHub's Community Standards} indicator.
We first extracted strategies used to document software features in textual artefacts and then strategies used to link the feature documentation with source code.
[Results]
We observed feature documentation in all studied projects in artefacts such as READMEs, wikis, and website resource files.
However, the features were often described in an unstructured way. 
Additionally, tracing techniques to connect feature documentation and source code were rarely used.
[Conclusions]
Our results suggest a lacking (or a low-prioritised) feature documentation in open-source projects, little use of normalised structures, and a rare explicit referencing to source code.
As a result, product feature traceability is likely to be very limited, and maintainability to suffer over time.
\end{abstract}


\section{Introduction \& Background}
\label{sec:introduction}

Social coding platforms such as GitHub often represent the main collaboration enabler for \gls*{oss} projects.
Contributors use GitHub features like its issue tracker, wiki, and pull requests to create connections between just-in-time requirements \cite{Ernst-empiRE-2012} and corresponding source code implementation. 
As a result, requirements documentation might become fragmented  across individual artefacts. 
No single contributor is likely to fully understand all requirements and the corresponding implementation details. Newcomers might even struggle to get started and involved in the projects \cite{Steinmacher-IEEE_Software-2019}.
To mitigate this knowledge gap, some GitHub projects document the overall requirements as product features in artefacts such as README files and wikis.
However, research lacks an understanding of how and to what extent these product features are documented and linked to source code.

Previous work has found that newcomers face barriers when trying to contribute to \gls*{oss} projects because of imprecise documentation~\cite{Steinmacher-IEEE_Software-2019, Constantino-ICGSE-2020}.
Developers and users wish for complete and up-to-date documentation \cite{Aghajani-ICSE-2019}.
A careful and well maintained project documentation can avoid losing active contributors \cite{Constantino-ICGSE-2020}.
Concerning understanding \gls*{oss} documentation, previous work has analysed the general structure of GitHub READMEs \cite{Prana-EMSE-2019, Liu-IST-2022}, identified the knowledge types in \gls*{api} reference documentation \cite{Maalej-TSE-2013} and source code comments \cite{Pascarella-MSR-2017, Steidl-ICPC-2013}, extracted functional requirements from blogs \cite{Pagano:EMSE:2012}, and developed approaches to augment documentation with more insightful sentences taken from \textit{Stack Overflow} \cite{Treude-ICSE-2016, Rahman-SCAM-2015}.
However, to the best of our knowledge, no previous work has explored the high-level product feature documentation across textual and source code artefacts in GitHub projects.

As a first step towards studying documentation strategies for product features, we explored 25 popular GitHub projects of different programming languages in a qualitative content analysis.
Our research goals were to understand:
\begin{description}
    \item[\textbf{RQ1a}] Which textual artefacts do GitHub projects use to document the product features?
    \item[\textbf{RQ1b}] What descriptive elements do GitHub projects use to document product features in textual artefacts?
    \item[\textbf{RQ2a}] What strategies do GitHub projects use to link textual product features to the source code?
    \item[\textbf{RQ2b}] What strategies do GitHub projects use to link source code to the textual product features?
\end{description}

We identified six artefact types that document product features, including multi-level READMEs, repository descriptions, and website resources.
We extracted nine descriptive element types to document product features, including the use of tables and images.
Finally, we observed six strategies to link features to source code and four to link code to the features.
However, a large portion of documented features did not offer a link to the corresponding implementation and was also not linked in the implementation, indicating lacking traceability.
Our exploratory study of documentation strategies can help project contributors  improve the process and structure of their feature documentation. 
We also draw attention to the need for feature documentation guidelines and tracing tools.

\section{Methodology}
\label{sec:method}

\begin{table}
\caption{Studied GitHub projects}
\begin{center}
\begin{tabular}{p{0.13\textwidth-2\tabcolsep} p{0.35\textwidth-2\tabcolsep}}
    \toprule
    \textbf{Language} & \textbf{Projects} \\
    \midrule
    C & HandBrake$^{\bigtriangleup}$ \\
    C++ & terminal$^{\bigtriangleup}$, winget-cli$^{\bigtriangleup}$ \\
    C\# & maui$^{\dagger}$, ReactiveUI$^{\dagger}$ \\
    Java & PhotoView$^{\circ}$, seata$^{\circ}$ \\
    JavaScript & covid19india-react$^{\bigtriangleup}$, create-react-app$^{\dagger}$, material$^{\circ}$, nightwatch$^{\circ}$, node-red$^{\bigtriangleup}$, spectacle$^{\bigtriangleup}$, swiper$^{\circ}$ \\
    Objective-C & IQKeyboardManager$^{\circ}$ \\
    Python & django-rest-framework$^{\dagger}$, mypy$^{\circ}$, numpy-ml$^{\circ}$, ParlAI$^{\circ}$, qutebrowser$^{\bigtriangleup}$ \\
    Swift & SwiftEntryKit$^{\circ}$ \\
    TypeScript & bit$^{\circ}$, kibana$^{\bigtriangleup}$, supabase$^{\bigtriangleup}$, webdriverio$^{\circ}$ \\
    \bottomrule
    \multicolumn{2}{l}{$^{\bigtriangleup}$End-user Application, $^{\circ}$Developer Library, $^{\dagger}$Application Framework}
\end{tabular}
\label{tab:projects}
\end{center}
\end{table}

We conducted an in-depth exploratory qualitative content analysis of 25 popular GitHub projects that provided the documentation artefacts recommended by GitHub's \textit{Community Standards}\footnote{e.g., \url{https://github.com/dotnet/maui/community}} indicator.
We applied our sampling criteria to Dabic et al.'s GitHub dataset~\cite{Dabic-MSR-2021} to first obtain ten projects for a pre-study to understand the data and then to create our sample of 25 projects.
We addressed RQ1 with an open-coding content analysis of textual artefacts, extracting product feature documentation strategies.
We then addressed RQ2 in two parts: (a) an additional open-coding content analysis of textual artefacts looking for traceability strategies to link the textual product features to the source code; and (b) an algorithmic and open-coding content analysis of source code files, looking for traceability strategies to link the source code to textual product features.
The first author performed the entire manual content analysis, assisted by regular discussions with the other authors throughout the process.

\textbf{Sampling.}
To form the dataset for our manual, time-intensive in-depth analyses, we sampled popular GitHub projects that most likely contained well-documented product features and traceability strategies.
We used the \textit{GitHub Search} dataset by Dabic et al.~\cite{Dabic-MSR-2021} as a starting point to mitigate the limitations of GitHub's \gls*{api}.
The dataset was released in early 2021 and contains the metadata of 735,669 projects of ten primarily used programming languages.
To gather up-to-date metadata from the projects, we queried GitHub's GraphQL \gls*{api}.
Our criteria for popular projects were $>=$ 5,000 stars and not marked as a fork, archived, disabled, locked, or private.
We also required that the last commit to the main branch was not older than one year ($>$ June 3, 2020).
These sampling criteria led to a sample of 2,537 projects.

We then conducted a pre-study with one project of each of the ten programming languages.
We learned that some popular projects use GitHub to primarily make their code public and not to grow a community of contributors (that would prioritise good publicly available documentation).
To address this, we used a \textit{Beautiful Soup}\footnote{\url{https://pypi.org/project/beautifulsoup4/}}-powered script to add the fulfilment of GitHub's Community Standards indicator as an additional sampling criterion for our study.
The indicator displays, at a glance, if a project provides important collaborative artefacts, like a README and a contribution guideline.
This reduced our target population to only 306 popular projects\footnote{None of the \textit{Kotlin}-based projects of the dataset fulfilled all selection criteria. So our final sample is limited to nine programming languages.}, each with -- at least seemingly -- high documentation quality.

The last step in the sampling process was to reduce our number of projects to a manageable number since our in-depth analyses requires ample time per project.
We applied a stratified random sampling strategy using the programming language as a characteristic to sort the projects into strata.
We sampled about seven to nine per cent of projects per stratum, with at least one project per language, resulting in the sample of 25 studied projects listed in Table \ref{tab:projects}.

\textbf{RQ1 Analysis.}
We conducted an in-depth exploratory qualitative content analysis on textual artefacts of each  sampled project.
Per project, we algorithmically extracted all textual files with the suffixes\textit{~.md,~.mdx,~.markdown,~.txt,~.asciidoc,} and~\textit{.rst} while analysing wiki pages directly via the GitHub website.
We did not analyse text from issues and pull requests since we consider them a record of project evolution and not a representation of the project's \textit{current state}.
The exploratory content analysis was guided by (but not limited to) a coding guide that we created during the pre-study.
We annotated product features and analysed the strategies used to document them.
We first analysed  READMEs, documentation directories, repositories descriptions, and wikis.
Afterwards, we considered the resources of external product websites, which 12 projects stored as text files within their repository.

\textbf{RQ2 Analysis.}
We conducted an in-depth exploratory qualitative content analysis on the textual artefacts (RQ2a) and source code files (RQ2b) in each GitHub project of Table \ref{tab:projects}.
For RQ2a, we analysed the same textual artefacts as for RQ1.
For RQ2b, we algorithmically extracted all source code files with a file suffix associated with the respective repository's primary programming language.
We then annotated strategies to create trace links from the textual product features to the source code (RQ2a) and from the source code to the textual product features (RQ2b).
Such a trace link can be, for example, the name of a feature-describing textual artefact.
Since the number of source code comments per repository is often high, we conducted two analysis iterations for RQ2b. 
The first iteration was a keyword search of the respective repository's comments to find ones related to the product features extracted for RQ1.
In the second iteration, we used the remaining time to analyse the remainder of the file from top to bottom.

For RQ1a, RQ1b, and RQ2a combined, each project was given a maximum of 120 minutes for extraction.
We time-boxed the analysis of RQ2b  to additional 120 minutes.
We report the results as means across all repositories.
Our replication package\footnote{\url{https://doi.org/10.5281/zenodo.6914643}} includes a snapshot of the detailed research data and all computational notebooks created for the analysis.

\section{Results}
\label{sec:results}

\subsection*{(RQ1a) Textual Artefacts that Document Product Features}
\label{sec:results_rq1a}

\begin{figure}[t]
\centerline{\includegraphics[width=\columnwidth]{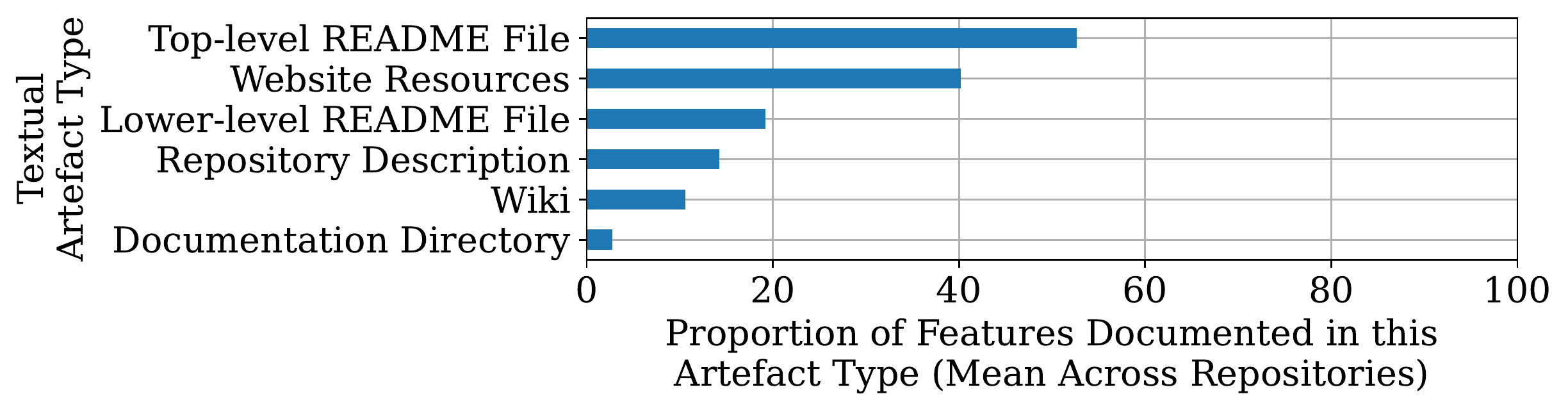}}
\caption{Textual artefacts that document product features}
\label{fig:textual_artefacts}
\end{figure}

Overall, we identified six types of textual artefacts used to document product features.
Figure~\ref{fig:textual_artefacts} shows the six types, ranked by the proportion of product features that appear in each of them, averaged (mean) across all 25 projects.
\textit{Top-level README files} were the most common, with $\approx$53\% of all product features appearing here.
\textit{Website resources}, containing $\approx$40\% of all product features, are subdirectories inside the GitHub projects containing the files that generate the project website.
\textit{Lower-level READMEs}, containing $\approx$19\%, are the files that maintainers place in subdirectories to explain parts of the project in more detail.
For example, \textit{create-react-app} and \textit{numpy-ml} used a README in the top-level directory to provide an overview of the repository content and applied lower-level READMEs extensively to detail the subdirectory contents.
The short \textit{repository descriptions} contained $\approx$14\%, while GitHub \textit{wikis} contained $\approx$11\% of product features. 
Finally, \textit{documentation directories} contained $\approx$3\%.
Documentation directories include multiple text files and could be unpopular because the GitHub website only renders the content of files called \textit{README}.
Therefore, contributors likely prefer to place all documentation in README files or use GitHub's wiki if multiple documentation pages are necessary.

The total number of software product features described in a GitHub project's documentation highly differed in our sample.
While we found 44 features in the \textit{maui} project, we extracted only one from the very sparse documentation of \textit{covid19india-react}.
A single product feature could occur in multiple artefacts, in which case we assigned the codes of all relevant artefact types to the feature.
We observed that the maintainers often used multiple artefact types to describe a single feature in varying levels of detail: for example, the repository description for a high-level description and the wiki for a more detailed explanation.
Such a double assignment of artefact types to one feature occurred 73 times; triple assignments happened 15 times. 
For 339 features, we only found one artefact type.

\subsection*{(RQ1b) Descriptive Elements to Document Product Features}
\label{sec:results_rq1b}

\begin{figure}[t]
\centerline{\includegraphics[width=\columnwidth]{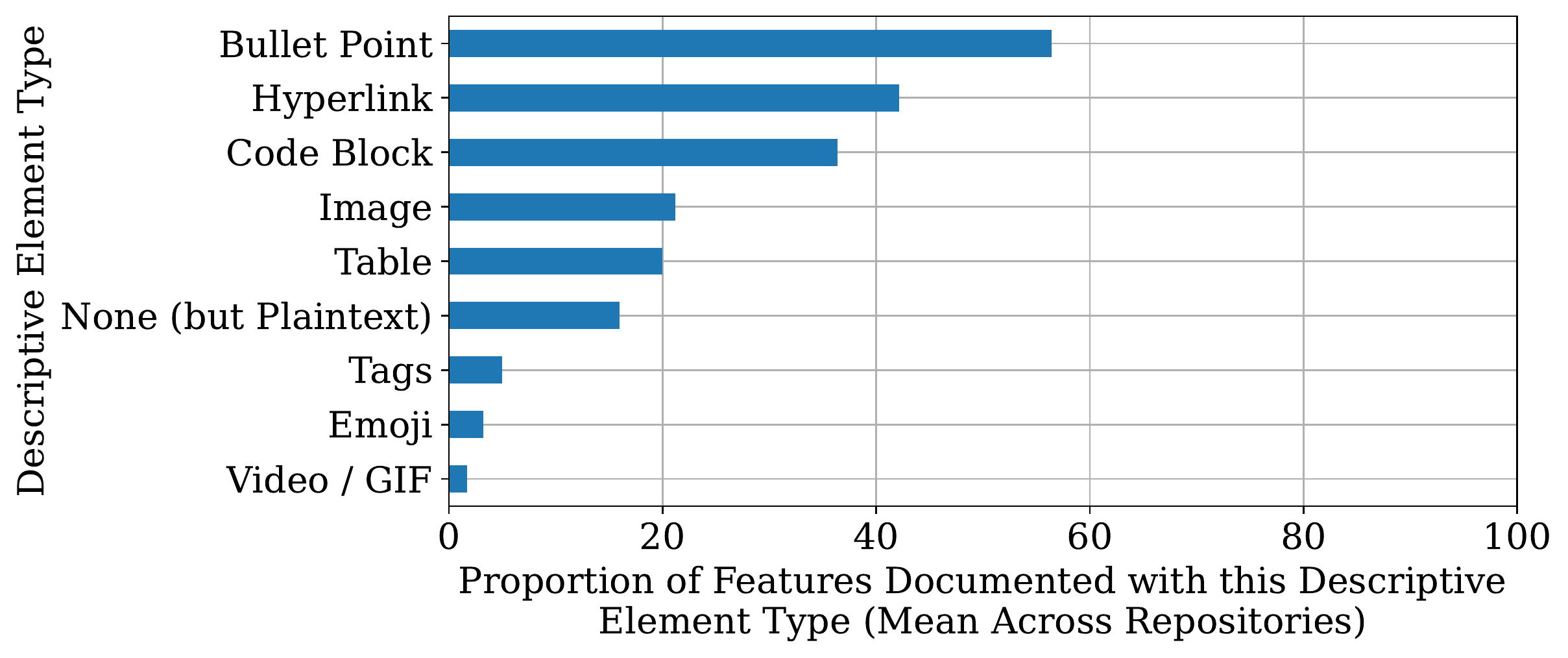}}
\caption{Descriptive elements  to document product features}
\label{fig:desc_elements}
\end{figure}

We extracted nine types of descriptive elements used to document product features in GitHub projects.
Eight of these elements were used in addition to simple plaintext, for example, to highlight the features in a README file.
Figure~\ref{fig:desc_elements} shows the nine descriptive element types, ranked by the proportion of product features that incorporated them (not mutually exclusive), averaged (mean) across all 25 projects.

The projects used up to five different types to describe a single feature across multiple artefacts; the median was two.
\textit{Bullet point} was the most-used type, appearing in $\approx$56\% of product features.
This suggests the desire for brevity and simple structure when detailing product features.
\textit{Hyperlinks} were the second most-used type, $\approx$42\%, outlining the desire to link to additional resources.
Most often, these were links to external websites. 
However, some product features also linked to internal subdirectories, text files, or source code to refer to further documentation or the source code implementation.
\textit{Markdown}-based \textit{code blocks} were the third most-used type at $\approx$36\%.
These code blocks formally described how to use certain features, similar to code examples in \gls*{api} documentation.
\textit{Images} were also fairly common ($\approx$21\%) and often used to present the capabilities of the product features.
\textit{Tables} were used with  $\approx$20\% of the features, mainly to structure the feature presentation and add additional short descriptions.
There were $\approx$16\% of the product features that had no particular description strategy beyond plaintext.

Finally, we also observed several rather rare descriptive elements. \textit{Tags} ($\approx$5\%), \textit{emojis} ($\approx$3\%), and \textit{videos / GIFs} ($\approx$2\%) account for the least-used feature description types.
Tags were used, for example, by the \textit{kibana} project to highlight the components related to a feature.
\textit{SwiftEntryKit} used \textit{GIF} files to visualise its interactive features.
\textit{Maui} used emojis to communicate the implementation status of its features without lengthy textual descriptions; while a green tick referred to fully implemented features, yellow and red emojis indicated that the implementation is still ongoing or not started yet.

\subsection*{(RQ2a) Strategies to Link Product Features to Source Code}
\label{sec:results_rq2a}

\begin{figure}[t]
\centerline{\includegraphics[width=\columnwidth]{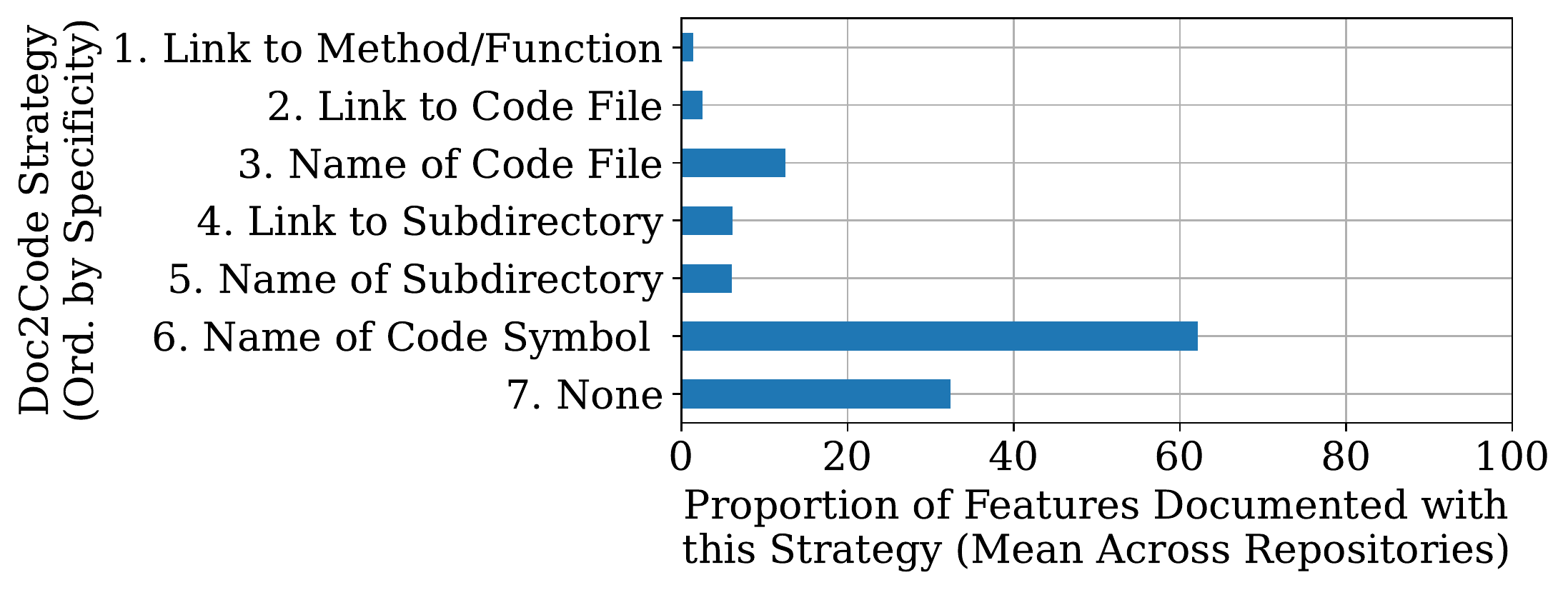}}
\caption{Strategies to link product features to source code}
\label{fig:doc_2_code}
\end{figure}

We extracted seven strategies to link product features to source code.
Figure~\ref{fig:doc_2_code} shows the strategies plotted with the proportion of product features that incorporate each of them, averaged (mean) across all 25 projects.
The documentation of a feature can use multiple strategies (not mutually exclusive).

The strategies are ranked by their specificity, meaning that the top strategy, \textit{link to method/function}, was the most direct way to trace the source code; however, occurring in only $\approx$1\% of documented product features.
Such a hyperlink referred directly to a line in a code file. 
This strategy introduces high maintenance costs since the link has to be updated if the line of code does not correspond to the method/function anymore after a code change.
The \textit{link to code file} strategy causes less maintenance. However, it was rarely used ($\approx$3\%).
For example, \textit{ParlAI} used these links to refer to more generally usable auxiliary files.
In comparison, just mentioning the \textit{name of code file} was more prevalent in our sample ($\approx$13\%).
The strategies \textit{link to subdirectory} ($\approx$6\%) and \textit{name of subdirectory} ($\approx$6\%) were usually used by the projects to give an overview of the product components distributed across multiple subdirectories of the repository. 
Often, these subdirectories contained additional READMEs to detail the component; an example is the project \textit{webdriverio}.
The most often used but less specific strategy was \textit{name of code symbol} ($\approx$62\%).
For this strategy, the projects presented plaintext or Markdown-based code blocks to enrich the feature description with the feature usage.
For example, \textit{supabase} provided code blocks in its documentation to explain the product's \gls*{api}.
The shortcoming of the sole use of this strategy is that it is mainly directed to the product users.
(Potential) project contributors would have to manually search the repository with the specific code symbol to find the corresponding implementation.
Nevertheless, this is still more specific than 
the \textit{none} strategy, meaning that the description of a product feature neither presented information about how to use the feature nor where to find its corresponding source code.
This strategy was applied to $\approx$32\% of the features, indicating a lack of tracing strategies for product feature documentation.

\subsection*{(RQ2b) Strategies to Link Source Code to Product Features}
\label{sec:results_rq2b}

\begin{figure}[t]
\centerline{\includegraphics[width=\columnwidth]{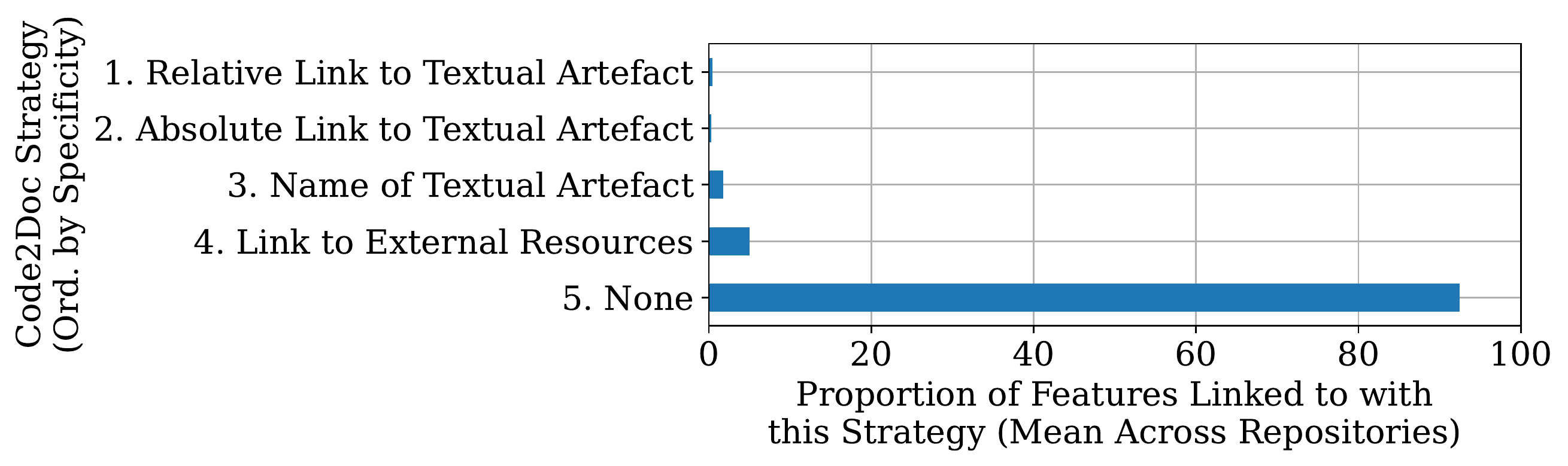}}
\caption{Strategies to link source code to product features}
\label{fig:code_2_doc}
\end{figure}

We extracted five strategies to link source code to the product feature documentation.
Figure \ref{fig:code_2_doc} lists these strategies plotted with the proportion of product features linked to with these strategies, averaged (mean) across all 25 projects and ranked by specificity.
The strategies are not mutually exclusive, meaning multiple code comments can apply different strategies to refer to a single product feature.

\textit{Relative} ($\approx$0.4\%) and \textit{absolute links to textual artefacts} ($\approx$0.3\%) as well as mentioning the \textit{name of a textual artefact} ($\approx$2\%) were rarely applied tracing strategies.
Relative links are more advantageous than absolute ones since they can be used in any environment to navigate from the source code to the documentation, including the local repository on a developer's computer.
The projects \textit{material}, \textit{kibana}, and \textit{ParlAI} used hyperlinks or names of textual artefacts to refer to configurations used in their source code, which the corresponding feature documentations described in more detail.
\textit{Link to external resources} was a slightly more popular strategy to refer to product features ($\approx$5\%). 
We included this strategy since these links most often referred to the product's externally hosted website.
In other cases (for example, \textit{ParlAI}), the links referred to scientific papers that provide specifics about a feature.
Nevertheless, for the majority of product features ($\approx$92\%), we could not find tracing strategies more sophisticated than the sole textual similarity between the code comments and the product feature documentation, indicating a lack of tracing techniques applied to refer from source code to feature documentation. 

\section{Discussion}
\label{sec:discussion}

The studied GitHub projects often used combinations of textual artefact types to document product features.
For example, the \textit{terminal} project  used top- and lower-level README files and a dedicated documentation directory to describe the features in varying levels of detail.
A combination of multi-level READMEs can help project contributors, especially newcomers, understand the code base's components.
On the contrary, other projects heavily used their external website to describe product features and stored the related website resources in the GitHub repository.
These resources are not a \textquote{native} Git documentation component, making them hard to consume on GitHub and limiting the usage of tracing strategies, like relative links between documentation and source code.
However, the projects could use a hybrid version of website resources and documentation directories:
They could enrich the website subdirectory with a README that explains the website resources' structure and provides hyperlinks to each resource file.
Readers could then browse the website resources more conveniently on GitHub without switching to the website.

We saw five relatively often and three less often used descriptive element types applied to enhance the plaintext description of product features.
Their usage highly differed across the GitHub projects and textual artefact types, implying that there is currently no consensus about specific strategies for documenting product features.
We assess that all of the observed descriptive element types can be beneficial in locating and understanding the feature documentation.
Further tools could motivate and facilitate their use in \gls*{oss} documentation.

Our analysis shows that -- unlike low-level tracing in pull requests and GitHub issues, the investigated projects barely use tracing techniques to connect the high-level functional knowledge about product features with the corresponding code implementation and components.
The use of more sophisticated tracing strategies can lead to a better understanding of the software components \cite{Kasauli-EASE-2020}, more efficient software maintenance and testing, and an improved requirements assessment \cite{Cleland-Huang-Future_SE-2014, Maeder-ICSM-2012}.
A simple approach to reach a better connection from documentation to code would be a section in the top-level README file that describes the structure of the project, its subdirectories, and software components.
In our sample, only \textit{terminal} provided such a section.

Besides, relative hyperlinks could help documentation readers efficiently navigate between product feature documentation and code.
This could reduce their need to search for specific code files or sections and help contributing newcomers get to know the project. 
If multiple code files correspond to one feature, the documentation could list multiple links next to the feature in a table-like structure or refer to the subdirectory containing the relevant code files.
However, a previous analysis of \gls*{oss} projects found that hyperlinks in source code comments are rarely updated elements since about 10\% of them led to non-available webpages \cite{Hata-ICSE-2019}.
This also stresses the need for tools that check the up-to-dateness of the links.

\section{Threats to Validity}
\label{sec:threats}

This \textit{qualitative} content analysis is a preliminary study to \textit{explore} currently used feature documentation strategies in GitHub.
The manual coding process was performed by the first author only.
We mitigated this threat to validity by running a pre-study and discussing all parts of the analysis approach, the findings, and the implications among all authors.
Moreover, the first author conducted a second coding iteration for the entire sample and research questions to mitigate annotation mistakes.

Our time-boxing resulted in an unfinished analysis of website resources for seven repositories and an unfinished second iteration of comment analysis for six projects.
We consider this acceptable; website resources are not the primary documentation means on GitHub, and we finished the highly-effective keyword-based first iteration of the comment analysis.

The target population consisted of 306 repositories from which we randomly sampled 25 for a time-intensive in-depth analysis.
Hence, the sample is likely not highly representative of the population. 
We consider the results as a basis for upcoming, more representative studies that could focus on specific aspects or artefacts of our broad analysis.

\section{Conclusions and Future Work}
\label{sec:future_work}

Our qualitative content analysis of 25 popular, potentially well-documented GitHub repositories revealed a broad and rather fragmented spectrum of strategies to document product features. 
We found six types of textual artefacts that listed features and nine types of descriptive elements presented these features.
Trace links between the analysed textual artefacts and source code were rare, which reveals an improvement potential for the software evolution as well as comprehension.

We presume that if the documentation artefacts in popular software projects are more interwoven with the source code, developers of aspiring projects will start to take care more and earlier about maintaining and enhancing the documentation.
Our analysis opens questions concerning the project contributors' rationale for their documentation decisions and their opinion regarding the improvement potential we identified.
Hence, future work can consider interviews and surveys with GitHub contributors and an analysis of available best practices in scientific and grey literature to triangulate findings from three perspectives: the project artefacts, the practitioners, and the best practices.
The results should not only establish and scrutinise documentation strategies in open-source settings, but also set the foundation for tools extending previous work \cite{Knauss-RE-2018} to facilitate documentation and overall collaboration.

\section*{Acknowledgement}
This work is partly funded by DASHH Data Science in Hamburg, the Helmholtz Graduate School for the Structure of Matter.

\bibliographystyle{IEEEtran}
\bibliography{ref}

\end{document}